\documentstyle[aps,pre,preprint,tighten]{revtex}
\input prepictex \input pictex \input postpictex
\begin{document}
\tolerance=10000
\draft
\title{
Ordinary, extraordinary, and normal surface transitions:\\
extraordinary-normal equivalence and simple\\
explanation of $|T-T_c|^{2-\alpha}$ singularities
}
\author{Theodore W.\ Burkhardt}
\address{
Department of Physics, Temple University, Philadelphia, PA 19122, USA
}
\author{H.\ W.\ Diehl}
\address{
Fachbereich Physik, Universit\"at - Gesamthochschule - Essen,\\
D-45117 Essen, Federal Republic of Germany
}
\date{\today}
\maketitle

\begin{abstract}
With simple, exact arguments we show that the surface magnetization
$m_1$ at the extraordinary and normal transitions and the surface
energy density $\epsilon_1$ at the ordinary, extraordinary, and normal
transitions of semi-infinite $d$-dimensional Ising systems have
leading thermal singularities $B_\pm |t|^{2-\alpha}$, with the same
critical exponent and amplitude ratio as the bulk free energy
$f_b(t,0)$. The derivation is carried out in
three steps: (i) By tracing out the surface
spins, the semi-infinite Ising
model with supercritical surface enhancement $g$ and vanishing
surface magnetic field $h_1$ is mapped exactly onto a
semi-infinite Ising model with subcritical surface enhancement, a
nonzero surface field, and irrelevant additional surface
interactions. This establishes the equivalence of the
extraordinary ($h_1=0,\,g>0$) and normal ($h_1\neq 0,\,g<0$)
transitions. (ii) The magnetization $m_1$ at the interface of an
infinite system with uniform temperature $t$ and a nonzero magnetic
field $h$ in the half-space $z>0$ only is shown to be proportional to
$f_b(t,0)-f_b(t,h)$. (iii) The energy density $\epsilon_1$ at
the interface of an infinite system with temperatures $t_+$ and $t$
in the half-spaces $z>0$ and $z<0$ and no magnetic field is shown
to be proportional to $f_b(t,0)-f_b(t_+,0)$.
\end{abstract}

\narrowtext
\section{INTRODUCTION}

In this paper we consider the critical behavior of macroscopic systems
with surfaces, walls, or interfaces on approaching the bulk
critical point \cite{bdl,hwd}. As is well known, the critical behavior
near boundaries normally differs from the bulk
behavior. The values of the bulk and surface critical exponents
characterizing thermodynamic singularities usually are different,
and the surface exponents are not generally expressible in terms of the
bulk exponents.

There are, however, a few remarkable cases in which local surface
quantities have thermal singularities of the same form $|t|^{2-\alpha}$
as the bulk free energy \cite{bm,dde,bc}. Here $t\sim T-T_c$, where
$T_c$ is the {\it bulk} critical temperature. Examples of such
quantities are (i) the surface order parameter $m_1$
at the extraordinary and normal transitions and
(ii) the surface energy density $\epsilon_1$ at the
ordinary, extraordinary, and normal transitions.

Our use of the terms \lq extraordinary\rq\ and \lq normal\rq\ requires
some explanation. The names ordinary, special, and extraordinary
were originally introduced \cite{lr} to distinguish the surface
transitions of the $d$-dimensional semi-infinite Ising model at the
bulk critical temperature {\it in the absence
of fields breaking the
$Z\!\!\!Z_2$ symmetry of the Hamiltonian.} Which surface transition
occurs depends on the enhancement $g$ of the surface interactions
\cite{nf}. The ordinary, special, and extraordinary transitions
correspond to $g<0$, $g=0$, and $g>0$, respectively. The extraordinary
transition originally referred to the transition from a
high-temperature phase with spontaneous surface order but no bulk order
to a low-temperature phase with both surface and bulk order.

Bray and Moore \cite{bm} formulated a
phenomenological theory of the
extraordinary transition based on

(a) the assumption that instead of being {\em spontaneously} ordered
in the high-temperature phase, the surface spins could just as well
be aligned by a surface magnetic field $h_1$, i.e., for $h_1\ne 0$
and arbitrary $g$ there should also be extraordinary behavior,

(b) a local free-energy scaling hypothesis.

\noindent Bray and Moore predicted that the leading non-analytic term
of the surface magnetization $m_1$ has the form $|t|^{2-\alpha}$.
Most authors accepted (somewhat
uncritically) assumption (a) and subsequently referred to both kinds of
transitions, $h_1=0,\,g>0$ and $h_1\neq 0,g<0$, as extraordinary
transitions.

The name extraordinary may be acceptable for systems with
$h_1=0,\thinspace g>0$, but it is inappropriate for systems
with $h_1\ne 0$ and $g<0$.
In contrast to magnetic systems, the Hamiltonian of fluid
systems at bulk coexistence is not normally $Z\!\!\!Z_2$ symmetric.
The surface field $h_1$ is
generically nonzero \cite{dc}. Hence the case
$h_1\ne 0$ and $g<0$ is actually quite normal for fluid
systems \cite{fu}. It applies, in particular, to the critical
adsorption of fluids \cite{fdg,lf,d,Law}. Following a suggestion by Fisher
\cite{mef}, we refer to the case $h_1\ne 0$ as {\it
normal} and reserve the name {\it extraordinary}
for $h_1=0,\,g>0$.

According to assumption (a) the normal and
extraordinary transitions should belong to the same surface
universality class. This is indeed the case. One source of evidence
is from field-theoretic
renormalization-group (RG) analysis in $4-\epsilon$
dimensions. One can show that the asymptotic critical
behavior at both transitions
is the same to any (finite) order of RG-improved perturbation
theory \cite{ds}.

In this paper we give an alternate, more direct proof of the
extraordinary-normal equivalence. This is presented in
Sec.\ II, where the semi-infinite Ising model
with supercritical surface enhancement is mapped exactly onto
a semi-infinite Ising model with $h_1\neq 0$ and subcritical surface
enhancement by tracing out the surface spins.

In Sec.\ III we show that the surface order parameter $m_1$ at the
extraordinary and normal transitions and the energy density
$\epsilon_1$ at the ordinary, extraordinary, and normal transitions
all have leading thermal singularities $B_\pm |t|^{2-\alpha}$, with
the same critical exponent and amplitude ratio
as the bulk free energy. Diehl et al.\ \cite{dde} first predicted
the $|t|^{2-\alpha}$ singularity in $\epsilon_1$ at the ordinary
transition as a consequence of a Ward identity. Later
Burkhardt and Cardy \cite{bc} presented simple,
model-independent arguments for $|t|^{2-\alpha}$ singularities
in $\epsilon_1$ at both the ordinary and normal transitions and in
$m_1$ at the normal transition. They also showed that conformal
invariance in two-dimensional semi-infinite systems implies
$|t|^{2-\alpha}$ surface singularities in a certain class of
densities. More recently it was claimed
\cite{tsallis,oo} that the leading
singularity in $m_1$ at the extraordinary transition does
not correspond to $|t|^{2-\alpha}$ but to a discontinuity
in $\partial_t m_1$. However, this has been refuted by a a detailed
field-theoretic study \cite{ds}.

The derivation in Sec.\ III is a variation of the approach
in \cite{bc}. The arguments
are particularly simple and elucidate the origin of
the bulk free-energy singularities in $m_1$ and $\epsilon_1$.

While we do not aim (nor claim) to control all
steps of our derivations in
a mathematically rigorous fashion, the
conclusions should be exact.

\section{Equivalence of the extraordinary and normal transitions}

Consider a semi-infinite $d$-dimensional hypercubic lattice of
Ising spins. There are ferromagnetic interactions with
interaction constant $K=J/k_BT$ between all the nearest-neighbor
spins except the surface pairs, which have interaction
constant $K_1=J_1/k_BT$. Denoting the surface layer by ${\cal S}$,
the rest of the system by ${\cal S}_c$, and
the spins in ${\cal S}$ and ${\cal S}_c$ by $\sigma$ and $s$,
respectively, we write the Hamiltonian as
\begin{equation}
{\cal H}\{s, \sigma\}={\cal H}_d\{s\}+{\cal H}_{d-1}\{\sigma\}+{\cal
H}_{\text{int}}\{s, \sigma\}\;, \label{Ham1}
\end{equation}
\begin{equation}
{\cal H}_d\{s\}=-K \sum_{\langle {\bf i},\,{\bf j}\rangle
\not\subset {\cal S}} s_{\bf i}\, s_{\bf j}\;,\label{Ham2}
\end{equation}
\begin{equation}
{\cal H}_{d-1} \{\sigma\}=-K_1 \sum_{\langle{\bf i},\,{\bf  j}\rangle
\subset {\cal S}} \sigma_{\bf i}\, \sigma_{\bf j}\;,\label{Ham3}
\end{equation}
\begin{equation}
{\cal H}_{\text{int}} \{s, \sigma\} = -K
\sum_{\langle{\bf i},\,{\bf
j}\rangle\in {\cal S}\times{\cal S}_c}\sigma_{\bf i}\, s_{\bf j}\;,
\label{Ham4}
\end{equation}
where $\langle {\bf i},\,{\bf j}\rangle$ indicates a pair of
nearest-neighbor sites.

Being interested in phases with spontaneously
broken $Z\!\!\!Z_2$
symmetry, such as the surface-ordered, bulk-disordered phase,
we must ensure that the appropriate pure state is selected in
the thermodynamic limit. Following the conventional procedure,
we include magnetic-field terms
$-h\sum\sigma_{\bf i}-h\sum s_{\bf j}$
with $h>0$
in ${\cal H}\{s,\sigma\}$ and then
take the limit $h\to 0+$ after the thermodynamic limit.

The lower critical dimension for a surface-ordered, bulk-disordered
phase is 2. For $d>2$ the extraordinary transition
occurs on crossing the line ${ E}:\,K=K_c,\, K_1>K_1^{\text{sp}}$
in the $(K,K_1)$ plane. Here $K_c=K_c(d)$ is the (bulk)
critical value of $K$, and $K_1^{\text{sp}}(d)$ is the critical
value of $K_1$ at the special or multicritical point
\cite{bdl,hwd,bm}. (Since $K$ and $K_1$ are reduced coupling
constants, a typical approach $T \to T_c$ corresponds to $K
\to K_c$ at fixed $K_1/K$.) The surface
critical behavior across all points of ${E}$ should be the same. In
analyzing the behavior below, it will be convenient to take the
enhancement $g= (K_1-K_1^{\text{sp}})/K$ to be large.

For supercritical $K_1$ the surface spins are spontaneously ordered.
The intuitive idea underlying the extraordinary-normal equivalence
is that the ordered $\sigma$ spins subject the neighboring $s$ spins
to an effective magnetic field. To put this idea on a
firmer footing, we map the composite
system defined in Eqs. (\ref{Ham1})-(\ref{Ham4}) exactly on
a model for the $s$ spins alone by tracing out the $\sigma$ spins.

In terms of the effective Hamiltonian ${\cal H}_{\text{eff}} \{s\}$
defined by the partial trace
\begin{equation}
e^{-{\cal H}_{\text{eff}} \{s\}} = {\text{Tr}}_{\sigma}\,
e^{-{\cal H} \{s, \sigma \}}\;,
\label{Heff1}
\end{equation}
the partition function of the original system is given by
\begin{equation}
Z^{(d)} = {\text{Tr}}_s\, {\text{Tr}}_{\sigma}\, e^{-{\cal H}
\{s,\sigma \}}
= {\text{Tr}}_s\, e^{-{\cal H}_{\text{eff}} \{s\}}\;.\label{Z}
\end{equation}
Equations (\ref{Ham1})-(\ref{Ham4}) and (\ref{Z}) imply
\widetext
\begin{equation}
{\cal H}_{\text{eff}} \{s\} = F^{(d-1)}(K_1) + {\cal H}_d \{s\}
-\ln <e^{-{\cal H}_{{\rm int}} \{ s,\sigma\}} >_{{\cal H}_{d-1}}.
\label{Heff2}\end{equation}\narrowtext\noindent
The first two terms on the right-hand side are contributed by the
uncoupled subsystems of $\sigma$ and $s$ spins, respectively.
The first term is the free energy
$-\ln{\text Tr}_\sigma\exp[-{\cal H}_{d-1}\{\sigma\}]$ of
the surface subsystem, and the
second term is the same as in Eq. (\ref{Ham2}). Expanding the third
term in powers of $K$,
we obtain
\widetext
\begin{equation}
{\cal H}_{\text{eff}} \{s \} = F^{(d-1)} (K_1) +
{\cal H}_d \{s \}- h_1
\sum_{{\bf i}\in \tilde{{\cal S}}} s_{\bf i}
- \sum_{n=2}^{\infty} {1\over n!}\; K^n\,
\sum_{\{{\bf j}_{\alpha} \in\tilde{{\cal S}}\}}
G_{{\bf j}'_1 \ldots {\bf j}'_n} (K_1)\,s_{{\bf
j}_1} \ldots s_{{\bf j}_n}\label{Heff3}\;
\end{equation}
\begin{equation}
h_1=K\,m_b^{(d-1)}(K_1)=K\,G_{{\bf i}'}(K_1)\;,\qquad
G_{{\bf j}_1'\dots{\bf j}_n'}(K_1)=
\langle\sigma_{{\bf j}_1'}\dots\sigma_{{\bf j}_n'}
\rangle^C_{{\cal H}_{\rm d-1}}\label{Heff4}
\end{equation}
\narrowtext\noindent
Here $\tilde{\cal S}$ denotes the surface layer of $s$ spins
after elimination of the layer ${\cal S}$ of $\sigma$ spins. The
sites ${\bf j}'_\alpha\in {\cal S}$ and
${\bf j}_\alpha\in\tilde{\cal S}$ are nearest neighbors. The quantities
$m_b^{(d-1)}(K_1)$ and $G_{{\bf j}_1'\dots{\bf j}_n'}(K_1)$ are the
spontaneous magnetization and $n$th cumulant of the subsystem of
$\sigma$ spins with Hamiltonian ${\cal H}_{d-1}\{\sigma\}$
in Eq.(\ref{Ham3}).

The original semi-infinite system is exactly equivalent to the
semi-infinite system with one less layer and the Hamiltonian
${\cal H}_{\text{eff}}\{s\}$ in Eqs. (\ref{Heff3}) and (\ref{Heff4}).
As mentioned above, the universal features of the extraordinary
transition should be independent of $K_1$ for $K_1>K_1^{sp}(d)$.
We now study the transition
for $K_1>>K_c(d-1)>K_1^{sp}(d)\ $\cite{fp}. In this limit the
$d-1$ dimensional system of $\sigma$ spins is deep in the
low-temperature phase, and $m_b^{(d-1)}(K_1)=
1+O(e^{-4(d-1)K_1})$.

That the spontaneous ordering of the $\sigma$ spins gives rise
to a surface field $h_1$ in ${\cal H}_{\text{eff}}\{s\}$ is seen
quite explicitly in Eqs. (\ref{Heff3}) and (\ref{Heff4}).
The $n=2$ term of ${\cal H}_{\text{eff}}\{s\}$ increases the
nearest-neighbor coupling of spins $\langle{\bf i},{\bf j}\rangle$
in the surface $\tilde{\cal S}$ from the original
value $K$ to the larger value
$K+K_{\perp}^2\,G_{{\bf i}',{\bf j}'}(K_1)$.
Since the cumulant $G_{{\bf i}',{\bf j}'}(K_1)$ vanishes exponentially
as $K_1\to \infty$, the new nearest-neighbor coupling
is clearly subcritical for
sufficiently large $K_1$. The surface interactions
between more distant pairs and the multi-spin
interactions corresponding to $n\ge 3$ also vanish exponentially
for large $K_1$. (From the familiar low-temperature expansion
one sees that these interactions are bounded by increasingly large
powers of $\exp (-K_1)$ as either the separation or the number $n$ of
interacting spins increases.) All the terms with $n\geq 2$ in
Eq. (\ref{Heff3}) should be irrelevant \cite{remark}. On neglecting them,
${\cal H}_{\text{eff}}\{s\}$ reduces
to the standard nearest-neighbor Ising Hamiltonian with subcritical
surface couplings and a surface magnetic field $h_1\neq 0$. This
establishes the equivalence of the
extraordinary and normal transitions.

\section{Derivation of $|\lowercase{t}|^{2-\alpha}$ singularities}

\subsection{Critical behavior of $m_1$ at the extraordinary transition}

Next we now show that $m_1$ at the normal transition has a
leading thermal singularity $B_\pm |t|^{2-\alpha}$
with the same critical exponent and amplitude ratio as the bulk free
energy. Consider a continuum model
with a one-component order parameter $\phi ({\bf x_\parallel},z)$
defined on the $d$-dimensional block ${\cal B}$:
$0\leq x_\parallel ^i\leq M $ for $i=1,2,\dots,d-1$
and $ -L_2\leq z\leq L_1$.
Periodic and free boundary conditions are applied in
the ${\bf x}_{\parallel}$ and $z$ directions, respectively.
As shown in Fig. 1, the block
${\cal B}$ consists of an upper portion
${\cal B}_+$ with $0< z \leq L_1$, a
lower portion ${\cal B}_-$ with $-L_2 \le z < 0$,
and the interface ${\cal I}$ at $z=0$. The Hamiltonian
${\cal H}$ has a $Z\!\!\!Z_2$ symmetric part ${\cal H}_{\text {sym}}$
that describes bulk critical behavior in
the thermodynamic limit $L_1,\,L_2\,, M
\to \infty$. We choose the usual $\phi^4$ form \cite{hwd}
\begin{equation}
{\cal H}_{\text{sym}} \{\phi\} = \int_{\cal B} d^d x
\left[ \case{1}/{2}
\,(\nabla
\phi)^2 + \case{1}/{2}\,\tau \, \phi^2 + \case{1}/{4!}\, u \,
\phi^4 \right]\;.\label{Hsym}
\end{equation}
Denoting the symmetric part with critical values of $\tau$ and $u$ by
${\cal H}_{\text{sym}}^c$, we consider the complete Hamiltonian
\begin{equation}
{\cal H} \{\phi\} = {\cal H}_{\text {sym}}^c \{\phi\}
+t \,\int_{\cal B} d^d x\thinspace \phi^2
- h \,\int_{{\cal B}_+} d^d x \thinspace \phi\;. \label{Hpert}
\end{equation}
The second term corresponds to a uniform
temperature-like deviation from criticality and the
third term to a magnetic field in ${\cal B}_+$ only.

In the thermodynamic limit the total free energy $F$ of the system
has the asymptotic form
\widetext
\begin{equation}
F= A\,[L_1\,f_b (t,0) + L_2\, f_b(t,h)
+f_i(t,h) + f_s (t,h) + f_s(t,0)] + \ldots \;,\label{F}
\end{equation}
\narrowtext\noindent
where $A = M^{d-1}$ is the cross-sectional area of the system.
The first two terms are bulk contributions from ${\cal B}_-$ and
${\cal B}_+$. The next three terms are the free energies associated
with the interface at $z=0$ and the surfaces at $z=L_1$ and $-L_2$.

Now let us displace the interface upwards a small distance $\Delta L$,
increasing the height of ${\cal B}_+$ by $\Delta L$
and decreasing the height of ${\cal B}_-$  by $\Delta L$.
{}From Eq.\ (\ref{F}) the change in free energy is given by
\begin{equation}
\Delta F = A \,\Delta L\, [f_b (t,0)-f_b(t,h)] +
\Delta L \,o (A)\;.\label{DF1}
\end{equation}

The change in free energy can also be written in terms of
the corresponding change
\begin{equation}
\Delta {\cal H} = h \int_0^{\Delta L} dz
\int d^{d-1} x_{\parallel} \thinspace \phi
({\bf x}_{\parallel}, z)\;,\label{DH}
\end{equation}
of the Hamiltonian. Expressed in this way,
\begin{eqnarray}
\Delta F &=& - \ln \left \langle e^{- \Delta {\cal H}} \right \rangle
\nonumber\\
&=& A\, \Big( h \int_0^{\Delta L} dz\,\langle
\phi ( {\bf x}_\parallel, z )
\rangle + O[ ( \Delta L )^2 ] \Big)\;,\label{DF2}\
\end{eqnarray}
where translational invariance parallel to the interface has been used
in performing the ${\bf x}_{\parallel}$ integration.
Here $\langle \ldots
\rangle $ indicates a thermal average with the
Boltzmann factor $e^{-{\cal
H}}$ of the original (unperturbed) system. It is
understood that the ultraviolet
singularities of the theory have been
appropriately regulated (e.g.\ by a
large-momentum cutoff), so that all required
expressions are well-defined.

Since the (regularized) profile
$\langle \phi ({\bf x}_\parallel, z) \rangle$ varies
smoothly with $z$, the integral in Eq. (\ref{DF2}) can be replaced by
$\Delta L\,\langle \phi ({\bf x}_\parallel,0) \rangle$
in the limit $\Delta L
\to 0$. In the thermodynamic limit and the
limit $\Delta L \to 0$, Eqs. (\ref{DF1}) and (\ref{DF2})
imply
\begin{equation}
h\,m_1 = f_b (t, 0) - f_b (t, h)\;,\label{m1}
\end{equation}
where $m_1 = \langle \phi ({\bf x}_{\parallel}, 0) \rangle$ is the
magnetization at the interface. The term
$f_b (t, h \ne 0)$ is regular at
$t=0$, whereas $f_b (t, 0)$ has the
leading thermal singularity $B_{\pm}
|t|^{2-\alpha}$. Consequently $m_1$ has a leading
thermal singularity with the same universal critical
exponent $2-\alpha$
and amplitude ratio $B_+/ B_-$ as the bulk free energy.

Equation (\ref{m1}) determines the magnetization at
the interface of an infinite system
with a uniform temperature $t$ and a
magnetic field $h$ in the half-space $z>0$ only.
As $t$ is lowered, there is a transition
at $t=0$ from bulk disorder to bulk order
in the lower portion of the system. Since
the net effect of the upper portion
of the system is to provide an effective
magnetic field at the interface, the interface critical behavior
should belong to the same universality class as
the normal transition in the
semi-infinite geometry, which, as shown
in Sec.\,II, is equivalent to the extraordinary
transition. Since the interface critical behavior is determined
by Eq. (\ref{m1}), the surface magnetization
$m_1$ in both the normal and extraordinary transitions
should also have the leading thermal singularities with same thermal
critical exponent and amplitude ratio as the bulk free
energy.

\subsection{Critical behavior of $\epsilon_1$
at the ordinary and extraordinary transitions}

In a similar fashion the $|t |^{2-\alpha}$
singularity of the surface
energy density $\epsilon_1$ can be related
to the thermal singularity of
$f_b(t,0)$. Instead of the Hamiltonian (\ref{Hpert}) we consider
\begin{equation}
{\cal H} \{\phi\} = {\cal H}_{\text {sym}}^c \{\phi\}
+ t\,\int_{{\cal B}_-} d^d x \thinspace \phi ^2
+t_+\,\int_{{\cal B}_+} d^d x \thinspace \phi^2\;.
\end{equation}
Now there is no magnetic field, and the temperature deviations from
criticality $t$ and $t_+$ in ${\cal B}_-$ and ${\cal B}_+$,
respectively, are different.

Proceeding as in the previous section, we obtain
\begin{equation}
(t_+ -t)\,\epsilon_1
=f_b(t_+,0)-f_b(t,0)\;, \label{e1}
\end{equation}
where $\epsilon_1=\langle\phi^2({\bf x}_\parallel,0)\rangle$.
For fixed $t_+\neq 0$, $\epsilon_1$ clearly has
a leading thermal singularity $B_\pm |t|^{2-\alpha}$ with the
same critical exponent and amplitude ratio as $f_b(t,0)$.

Equation (\ref{e1}) gives the energy density
at the interface of an infinite system
with different temperatures $t_+$ and $t$
in the half-spaces $z>0$ and $z<0$, respectively,
and zero magnetic field.

For fixed $t_+>0$ the upper portion of
the system is in the high-temperature, disordered phase.
As $t$ is lowered, there is a transition at
$t=0$ from disorder to bulk order in the
lower portion of the system. Only for $t<0$
are the interface spins ordered, and
this order is driven by the bulk order.
For $t<t_+$ the net effect of the upper portion of the system
on the lower portion is to suppress $\phi$ near
the interface. All this suggests that the
interface transition with fixed $t_+>0$ belongs to
the universality class of the ordinary
transition in the semi-infinite geometry.

For fixed $t_+<0$, on the other hand,
the upper portion of the system is in the
low-temperature, ordered phase. As $t$ is lowered,
there is a transition at $t=0$
from a phase with interface order but no
bulk order in the lower portion to a
phase with both interface and bulk order.
The net effect of the upper portion of
the system is to provide an effective
magnetic field at the interface. Thus the
interface transition with fixed $t_+<0$ should belong to the
same universality class as the normal and
extraordinary transitions in the semi-infinite geometry.

Since the interface critical behavior for fixed $t_+$
is determined by Eq. (\ref{e1}), the surface energy
density $\epsilon_1$ at the ordinary, extraordinary,
and normal transitions should also have leading thermal singularities
with the same universal critical exponent
and amplitude ratio as the bulk
free energy.

\section{Concluding Remarks}

In summary we have derived bulk free energy
singularities in $\epsilon_1$ at the ordinary transition
and in $m_1$ and $\epsilon_1$ at the normal and extraordinary
transitions with simple, exact
arguments, without making the assumptions (a) and (b)
(see Sec. I) of Bray and Moore.

The predictions of leading thermal singularities
of the form $B_\pm |t|^{2-\alpha}$ in $\epsilon_1$ at
the ordinary transition and in $m_1$ at the
extraordinary and normal transitions check with
field theoretic results \cite{dde,ds} for the Ising model
in $d=4-\epsilon$. Comparable field theoretic results for
the singular behavior of $\epsilon_1$ at the
extraordinary transition are not yet available. As mentioned
above, $|t|^{2-\alpha}$ surface singularities in a certain class
of densities are also
implied by conformal invariance \cite{bc} in $d=2$.

As a check on our prediction that the $B_\pm |t|^{2-\alpha}$
singularities in surface quantities have the
same amplitude ratio $B_+/B_-$ as the bulk
free energy, we calculate the amplitude ratio for
$m_1$ at the extraordinary transition of the
Ising model from results of Diehl and Smock
\cite{ds} for the order-parameter profile in $d=4-\epsilon$.
Taking the distance from the surface to be small
in comparison with the bulk correlation length and
using Eqs.\ (47d), (48e), and (48f) in \cite{ds},
we obtain
\begin{equation}
{B_+\over B_-}={1\over 4}\,[1+O(\epsilon)]\;. \label{B+B-}
\end{equation}
This does indeed agree with the amplitude ratio \cite{dl}
\begin{equation}
{B_+\over B_-}=2^{\alpha-2}\,n\,\left[
1+\epsilon+O(\epsilon^2)\right]\;,\qquad
\alpha= {4-n\over 2(n+8)}\thinspace \epsilon + O(\epsilon^2)
\end{equation}
for the bulk free energy
of the $n$-vector model in the Ising case $n=1$.

Finally we point out that Ising spin variables are
not essential for the simple arguments of this paper.
The predictions should hold for a broad class of semi-infinite
systems with short range, ferromagnetic interactions and
second-order bulk transitions. Note, however, that the proof
of the extraordinary-normal equivalence in Sec. II does not
go through for systems with continuously broken symmetries,
due to the nonexponential decay of correlations
in the ordered phase.

\acknowledgements
We thank Erich Eisenriegler for useful discussions.
T.W.B. greatly appreciates the
hospitality of H. Wagner and coworkers
at the Universit\"at M\"unchen,
where part of this work was done, and the
support of the WE-Heraeus-Stiftung. H.W.D.\ acknowledges partial

support by Deutsche Forschungsgemeinschaft through Sonderforschungsbereich

237.

\begin{figure}
\beginpicture\thicklines
\setcoordinatesystem units <1.0cm,1.0cm> point at 0 0
\setplotarea x from -5.0 to 5.0, y from -10.0 to 10.0
\unitlength1mm

\put {\line(0,1){80}} [Bl] at 2.5 -4.0
\put {\line(0,1){80}} [Bl] at -2.5 -4.0

\put {\line(1,1){17.68}} [Bl] at 2.5 -4.0
\put {\line(1,1){17.68}} [Bl] at -2.5 4.0
\put {\line(1,1){17.68}} [Bl] at 2.5 4.0
\put {\line(1,0){50}} [Bl] at -2.5 -4.0
\put {\line(1,0){50}} [Bl] at -2.5 4.0
\put {\line(0,1){80}} [Bl] at 4.268 -2.232
\put {\line(1,0){50}} [Bl] at -0.732 5.768

\setdashes <2mm>
\plot -2.5 0.0 2.5 0.0 4.268 1.768 -0.732 1.768 -2.5 0.0 /
\thicklines
\setdots <1mm>
\plot -2.5 -4.0 -0.732 -2.232 4.268 -2.232 /
\plot -0.732 -2.232 -0.732 5.768 /

\put {$A$} at 0.8 5.0
\put {$h\ne 0 $} at 0.5 2.5
\put {$h=0$} at 0.5 -1.5
\put {${\cal B}_+$} at -3.4 1.8

\thinlines
\put {\line(2,1){9}} [Bl] at -3.05 1.8
\put {${\cal I}$} at -3.4 0.3
\put {\line(6,-1){10}} [Bl] at -3.05 0.3

\put {${\cal B}_-$} at -3.4 -1.8
\put {\line(2,-1){9}} [Bl] at -3.05 -1.8

\betweenarrows {$L_1$}

from 5.0 1.768 to 5.0 5.768
\betweenarrows {$L_2$}

from 5.0 -2.232 to 5.0 1.768
\thicklines
\setdots <1mm>
\setdashes <2mm>

 \hshade 0.0 -2.5 2.5  1.768 -0.732 4.268 /

\endpicture

\caption{Geometry of the system considered in Sec. IIIA.}
\end{figure}

\begin{references}
\bibitem{remark} {Note that the magnetic field term in
${\cal H}_{\text{eff}}\{s\}$ in Eqs. (\ref{Heff3}) and
(\ref{Heff4}) vanishes for
$K_1^{sp}(d)<K_1<K_c(d-1)$, even though the original system
has an extraordinary transition \cite{fp}.
For $K_1$ satisfying this inequality, the terms with $n\geq 2$ in
${\cal H}_{\text{eff}}\{s\}$ are not all irrelevant, since
they change the transition from ordinary
to extraordinary.}
\end{references}
\end{document}